\documentclass[11pt]{article}
\usepackage{amssymb,latexsym,amsmath}
\usepackage[english]{babel}
\usepackage{xcolor}
\usepackage{bm}
\usepackage{amsmath,amscd}
\usepackage{graphicx}
\usepackage{newtxtext}
\usepackage{newtxmath}
\usepackage{geometry}
\usepackage{hyperref}
\hypersetup{
    colorlinks = true,
    urlcolor   = blue,
    citecolor  = black,
}

\newcommand{\RomanNumeralCaps}[1]
\linenumbers

\title{Electric? Then it is geometric}

\author{\textbf{G. de Saxc\'e} \\
Univ. Lille, CNRS, Centrale Lille, UMR 9013 – LaMcube \\
Laboratoire de m\'ecanique multiphysique multi\'echelle, \\
F-59000, Lille, France, Email: gery.de-saxce@univ-lille.fr}

\begin{document}
\maketitle

\begin{abstract}
In this work, we revisit Kaluza-Klein theory from the perspective of the classification of elementary particles based on the coadjoint orbit method. We propose a symmetry group for which the electric charge is invariant and, on this basis, a cosmological scenario in which the three former spatial dimensions inflate quickly while the fifth one shrinks, leading to a 4D era where the particles correspond to the coadjoint orbits of this group. By this mechanism, the elementary particles can acquire electric charge as a by-product of the $4 + 1$ symmetry breaking of the Universe. By pullback over the space-time, we construct the non-Riemannian connection corresponding to this symmetry  group, allowing to recover conservation of the charge and the equation of motion with the Lorentz force. On this ground, we develop a five dimensional extension of the variational relativity allowing to deduce in the classical limit Maxwell's equation. 
\end{abstract}           


\vspace{0.2cm}






\section{Introduction}

First of all, what do we mean by geometric? A geometry is a group, it acts on objects and conserves invariants:
\begin{itemize}
    \item the affine group that conserves the ratio, the midpoint of Thales geometry.
    \item the Euclid group that conserves also distances,
    \item the Galilei group which also conserves the durations and the inertial motion. So this also concerns Physics. 
\end{itemize}
If you know the symmetry group, you can consistently develop a whole physical theory, with of course the help of other methods than those of group theory. Find the group and you will be able to explain this Physics by geometrization. 

If you ask: "electric?" we answer: "then, it is geometric".

Geometrizing electromagnetism was already done a century ago. It is Kaluza-Klein  theory, a promising approach but which suffers weaknesses.  Whatever the variant, there are flaws. It is not exactly electromagnetism, it is something else. In 1921,  Theodor Kaluza proposes in \cite{Kaluza 1915} to unify gravitation and electromagnetism, considering a 5D Universe equipped with a metric of signature $1 + 4$. Oskar Klein was Niels Bohr’s assistant, then he participated to the birth of Quantum Mechanics. He had the idea that the 5th dimension is curled up and overwhelmingly small, of the order of $10^{-31}$ cm \cite{Klein 1926, Klein 1926 Nature}. This explains why "we do not see the extra dimension" and allows to assume that all fields do not depend on the 5th coordinate. This is called the "cylinder condition".

The main flaw identified in the literature may be stated as follows:
there are 15 unknown, the 10 potentials of the gravitation, the 4 potentials of the electromagnetism and an extra potential called the dilaton. To find them, they are 15 field equations. Then where is the trap? The problem lies in the last equation: 
either the equation is used to determine the dilaton but has no physical meaning, 
or, as Oskar Klein, we put the dilaton to 1 but the equation is not satisfied. 
Until now, nobody succeeded in breaking this deadlock.

Author's opinion is that there is an even more annoying pitfall. In this theory, the electric charge would be the linear momentum along the 5th dimension. This is the starting point of this paper.  We ask the question: 

"As expected, according to the experience, is the electric charge independent of the choice of the reference frame, of the observer?" 

To answer it, we use the method of coadjoint orbits.

\vspace{0.3cm}

The paper is a trilogy: classification of the particles in Kaluza-Klein theory, pullback connection over the space-time, extended variational relativity. It is structured as follows.

In \textbf{Section 2}, we present the coadjoint orbit method allowing classification of the elementary particles in relativity and we apply it to the symmetry group of Kaluza-Klein 5D space, denoted $\hat{\mathbb{G}}_1$, which conserves the hyperbolic metric. We show that the electric charge is not preserved by the group, and therefore depends on the reference frame, contradicting the observations. To avert this paradox, we zoom in on the fifth dimension to reveal a new symmetry group denoted $\hat{\mathbb{G}}_0$ for which the charge is invariant. On this basis, we propose a cosmological scenario in which the elementary particles can acquire electric charge as a by-product of the 4 + 1 symmetry breaking of the Universe.

The second part of the trilogy is developed in  \textbf{Section 3}. Our aim is to construct the $\hat{\mathbb{G}}_0$-connections on the frame principal bundle. As the zoom out leads to a singularity in the fifth dimension, we endow the space-time with a pullback connection. We claim that the motion of a charged particle and the evolution of its charge are such that its linear 5-momentum is parallel-transported. The torsion free condition allows a demonstration that the charge is conserved along the trajectory and recovery of the Lorentz force.  

The trilogy ends with \textbf{Section 4} where we revisit Palatini variational relativity by adding to the ten potentials of the metric the four electromagnetic potentials. The generalized action depends on the matter and the connection representing both the gravitation and electromagnetic interactions. We discuss the field equations with coupling terms between the two interactions. At the Newtonian approximation, the coupling is weak and we recover the Maxwell equations.

\textbf{Section 5} is devoted to conclusions and perspectives. 

\vspace{0.3cm}

The aim of this paper is to present only the strength ideas of our new approach, leaving aside the too technical aspects or too lengthy developments. For more details, the reader is referred to the extended paper \cite{de Saxce 2024} where she or he can find extensive explanations of results which are presented here in an elliptical manner. 

\vspace{0.3cm}

This work was presented at the 66th Souriau Colloquium (formerly called “Colloque International de Théories Variationnelles”), Bastia (Corsica, France), 28 April-3 May 2024, and at the annual meeting of the GDR CNRS 2043 "Géometrie Différentielle et Mécanique", La Rochelle (Nouvelle-Aquitaine, France), 26-28 June 2024.

\begin{figure}[h!]
\centering
\includegraphics[scale=.60]{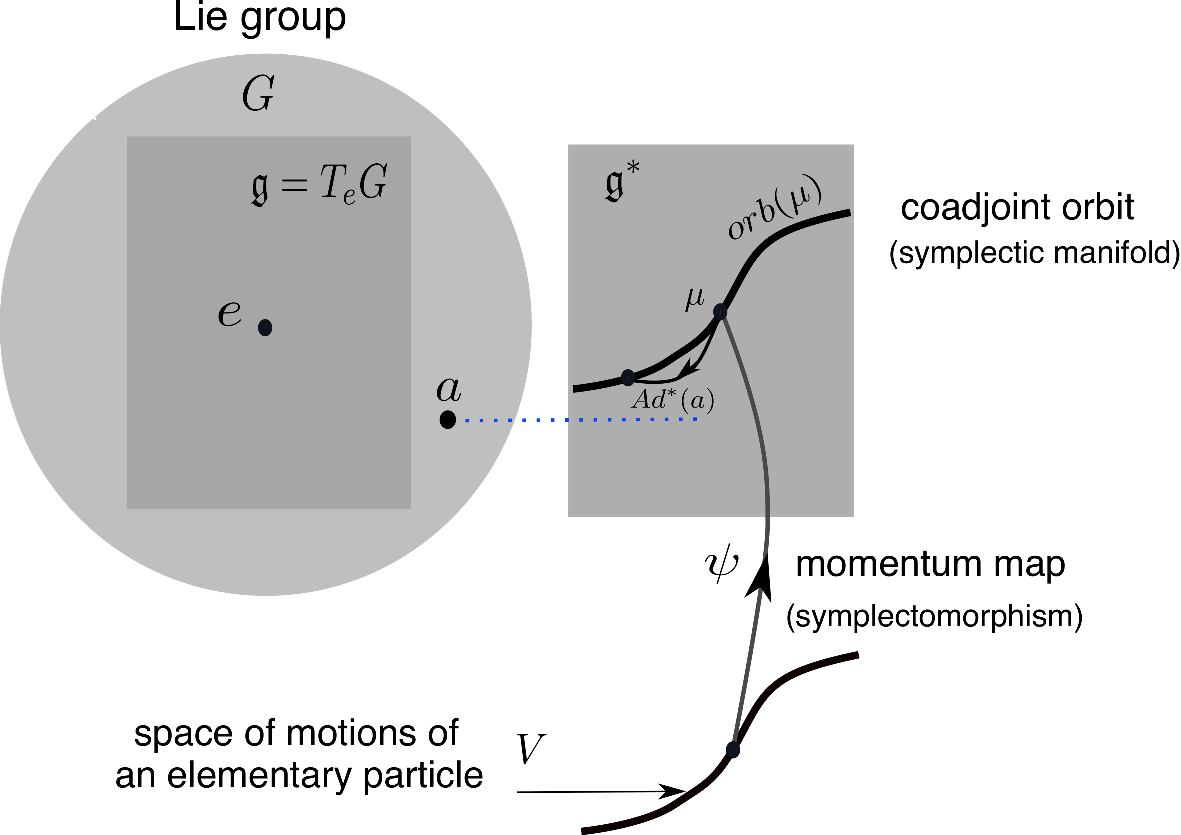}
\caption{Coadjoint orbit method}
\label{fig Coadjoint orbit method}
\end{figure}

\section{The coadjoint orbit method}

This method, the basis of the geometric quantization, was developed by Jean-Marie Souriau in his book "Structure of Dynamical Systems" \cite{SSD, SSDEng} to which the reader is referred for more details. The gist of the method is explained in Figure \ref{fig Coadjoint orbit method}. On the left, we have a symmetry group, the Lie group $G$ and its Lie algebra $\mathfrak{g}$, \textit{i.e.} the tangent space at the unity.
On the right, we have the dual $\mathfrak{g}^*$ of the Lie algebra on which act the elements of the group by the coadjoint representation. The orbit of a momentum $\mu$ has the structure of symplectic manifold.
The momentum map $\psi$ is a symplectomorphism (\textit{i.e.} an isomorphism of the symplectic structure) from the space of motions of an elementary particle into this orbit.
By classifying coadjoint orbits, one classifies the particles known to physicists, whether elementary such as quarks and leptons or composite such as hadrons.

\subsection{Example: Poincar\'e's group}

We start with an example to illustrate, the Poincaré's group $\mathbb{G} = \mathbb{R}^4 \rtimes \mathbb{SO} (1,3) $, that of the affine transformations $a = (C, P)$ of the 4D space conserving the Minkowski metric $G$ (of signature $1 + 3$), composed of a translation $C$ and a linear part $P$ which is unitary 
$$ P^* P = 1_{\mathbb{R}^4}
$$
and is called a Lorentz transformation. $\mathbb{G}$ is a Lie group of dimension $10$. The elements of the Lie algebra have an skew-adjoint linear part $\delta P$
$$ \delta a = (\delta C, \delta P) \in  \mathfrak{g}  
 \qquad   \Leftrightarrow \qquad 
 \delta P^* = - \delta P
$$
From the components of the momentum, the linear 4-momentum $\Pi$ and the angular 4-momentum $M$
$$ \mu \in \mathfrak{g}^*   
\qquad   \Leftrightarrow  \quad 
\mu = (\Pi, M) \quad \mbox{such that} \quad  M= - M^*
$$
we determine the polarization (or spin vector) 
$$ W = (*M) \, \Pi
$$
where $*$ is the Hodge operator.

\begin{figure}[h!]
\centering
\includegraphics[scale=.60]{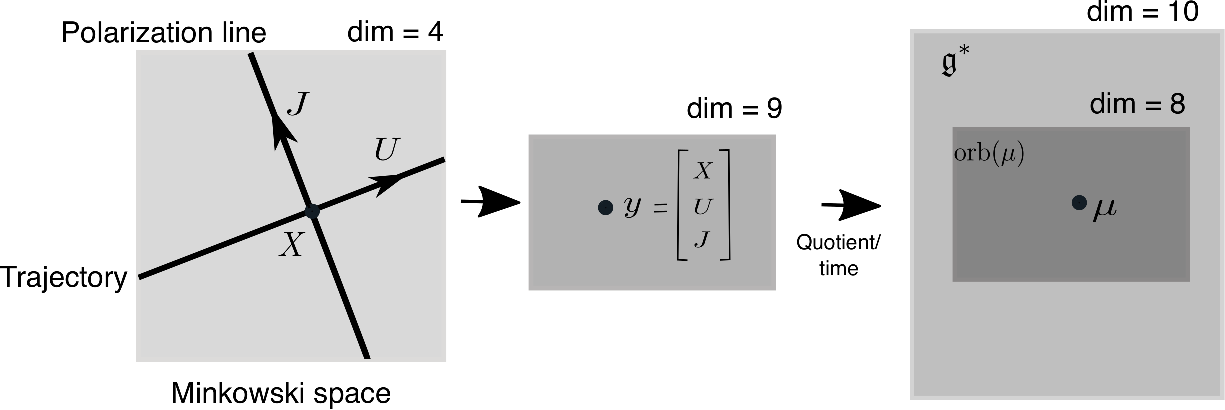}
\caption{Coadjoint orbit of a particle with spin for Poincar\'e group}
\label{fig Particle with spin Poincare}
\end{figure}

As represented in Figure \ref{fig Particle with spin Poincare}, every inertial motion of a particle with spin is defined by $y$, composed of the position in the space-time, X, called an event, the $4$-velocity $U$ (carried by the trajectory) and a $4$-vector of polarization $J$ (carried by a polarization line) for a total of $3 \times 4 = 12$ components.
$U$ and $J$ verifying 3 conditions of orthogonality and normalization, the space of $y$ is of dimension $12 - 3 = 9$.
By passing to the quotient by the time, the space of motions is of dimension $8$ and, by diffeomorphism, the coadjoint orbit is as well.
In the dual space of dimension $10$, that of the Poincaré's group, the orbit is defined by $10 - 8 = 2$ independent invariants which may be the rest mass
$$ m_0 = \sqrt{\Pi^* \Pi}
$$
and the spin
$$  s = \frac{\sqrt{- W^* W}}{\sqrt{\Pi^* \Pi}}
$$
There are other orbits of different dimensions characterizing the spinless particles and the massless particles.

\subsection{Charged elementary particles in Kaluza-Klein relativity (group $\hat{\mathbb{G}}_1$)}

Now we move to 5D. By convention, objects are topped with a hat. Our aim is to do the same exercise for a charged particle, with the Kaluza-Klein group $\hat{\mathbb{G}}_1 = \mathbb{R}^5 \rtimes \mathbb{SO} (1,4) $  of 5D affine transformations that conserve the metric of signature $1 + 4$. Their linear part is a $5 \times 5$ matrix broken into blocks
$$ \hat{C} = \left[ {{\begin{array}{cc}
       C \hfill  \\
       \xi  \hfill  \\
   \end{array} }} \right], \qquad
\hat{P} = 
\left[ {{\begin{array}{cc}
       P \hfill &  \beta \, P^{*-1} b  \hfill \\
       b^* \hfill &  \beta \hfill \\
   \end{array} }} \right]
$$
where 
\begin{eqnarray}
    b\in \mathbb{R}^4, \quad\qquad \beta = \sqrt{1 + b^* b} 
    \qquad\qquad \qquad\qquad\qquad\qquad \qquad\qquad
    \quad\quad \nonumber\\
P = P_L B, \qquad P_L \;\; \mbox{is a Lorentz transformation},
\qquad B = 1_{\mathbb{R}^4} + \frac{1}{\beta + 1} \, b \, b^* 
\end{eqnarray}
It is a Lie group of dimension 15. Following Souriau, the momenta of the elementary particles can be obtained by considering the group action on the coadjoint orbits. The momenta $\hat{\mu}$ of $\hat{\mathbb{G}}_1$ are of the form
$$ \hat{\mu} \in \hat{\mathfrak{g}}^*_1  \qquad \qquad   \Leftrightarrow \qquad \qquad \quad \hat{\mu} = (\hat{\Pi}, \hat{M}) \qquad \quad  \hat{M} \quad \mbox{such that} \quad \hat{M}^* = - \hat{M}
$$
We are interested in the component $\hat{\Pi}$  of the linear 5-momentum that we decompose  into the linear 4-momentum $\Pi\in \mathbb{R}^4$ and the electric charge $q$
$$ \hat{\Pi} = \left[ {{\begin{array}{cc}
       \Pi \hfill  \\
       q  \hfill  \\
   \end{array} }} \right]
$$
For a charged particle with spin, the coadjoint representation reads
$$ \Pi = P \, \Pi' + q' \, \beta \, P^{*-1} b, \qquad 
q = b^* \Pi + \beta \, q'
$$
and we see immediately that \textbf{the electric charge is not an invariant}. It depends on the choice of the reference frame, of the observer, in contradiction with experiment. Hence \textbf{the group} $\hat{\mathbb{G}}_1$ \textbf{of Kaluza-Klein is  not the symmetry group of the Universe today as we know it}. However, our opinion is that it must not be rejected \textit{a priori}. We just need to identify the Physics that admits this symmetry groop. But before that, \textbf{our goal now is to find a symmetry group for the Physics today}. 

\subsection{Charged elementary particles today (group $\hat{\mathbb{G}}_0$)}

As the scale of the Universe along the 5th dimension is overwhelmingly small, we zoom in
$$ \hat{X}'^5 = \hat{X}^5 / \omega
$$
in such a way that what was of the order of $10^{-31}$ cm is in the new coordinate of the order of $1$. 

When the small parameter $\omega$ --the cylinder radius-- tends to zero, the metric degenerates. In the new coordinates, it is represented in a basis by the $5 \times 5$ matrix $\hat{G}'$ which tends to $\hat{G}_0$ representing a bilinear form $\hat{\mathbb{G}}_0$ (but not a metric) 
$$      \hat{G}' = 
\left[ {{\begin{array}{cc}
       G \hfill &  0 \hfill \\
       0 \hfill &  - \omega^2 \hfill \\
   \end{array} }} \right] \rightarrow 
   \hat{G}_0 = \left[ {{\begin{array}{cc}
       G \hfill &  0 \hfill \\
       0 \hfill &  0 \hfill \\
   \end{array} }} \right]
$$
and $G'^{-1}$ tends to something proportional to the tensor product by itself of the 5-column $\Omega_0$ representing a vector $\hat{\bm{\Omega}}_0$
$$ \hat{G}'^{-1} \rightarrow
   - \omega^{-2} \, 
   \left[ {{\begin{array}{cc}
       0 \hfill  \\
       1  \hfill  \\
   \end{array} }} \right] \otimes 
   \left[ {{\begin{array}{cc}
       0 \hfill  \\
       1  \hfill  \\
   \end{array} }} \right] 
$$
The set of affine transformations $\hat{a} = (\hat{C}, \hat{P})$ of $\mathbb{R}^5$ that conserve the components of $\hat{\bm{G}_0}$ and $\hat{\bm{\Omega}}_0$
$$  \hat{G}_0 = \left[ {{\begin{array}{cc}
       G \hfill &  0 \hfill \\
       0 \hfill &  0 \hfill \\
   \end{array} }} \right], \qquad
    \hat{\Omega}_0 =
    \left[ {{\begin{array}{cc}
       0 \hfill  \\
       1  \hfill  \\
   \end{array} }} \right] 
$$
is a Lie group $\hat{\mathbb{G}}_0$ of dimension 15. Its elements $\hat{a}$  are such that their linear part is of the form
$$ \hat{P} = 
\left[ {{\begin{array}{cc}
       P_L \hfill &  0  \hfill \\
       b^* \hfill &  1 \hfill \\
   \end{array} }} \right]
$$
where $P_L$ is a Lorentz transformation. The momentum comprises, in addition to $\Pi, q$ and $M$, a 4-vector $Q$
$$ \hat{\mu} \in \hat{\mathfrak{g}}^*_0  \quad  \Leftrightarrow \quad \hat{\mu} = ((\Pi, q), (M, Q)) \quad  M^* = - M, \quad Q \in \mathbb{R}^4
$$
For an orbit of a charged particle with spin, the coadjoint representation reads
$$  q = q', \quad \Pi = P \, (\Pi' - q' \, b), \quad Q = P \, Q' + q' C 
$$
$$ M = P \, M' P^* + C \, (P \, (\Pi' - q \, b))^* - (P \, (\Pi' - q \, b)) \, C^*
    + (P \, b) (P \, Q')^* - (P \, Q') (P \, b)^*
$$
and we immediately see that --as expected-- \textbf{the electric charge is invariant}. If we calculate the dimension of the orbit, we find that there are 3 independent invariants taken to be the rest mass, the spin and the electric charge
$$  q, \qquad
   m_0 = \sqrt{\Pi^* \Pi}, \qquad 
   s = \frac{\sqrt{- W^* W}}{\sqrt{\Pi^* \Pi}}
$$

\subsection{A cosmological scenario for the evolution of elementary particle structure}

 \begin{figure}[h!]
\centering
\includegraphics[scale=.35]{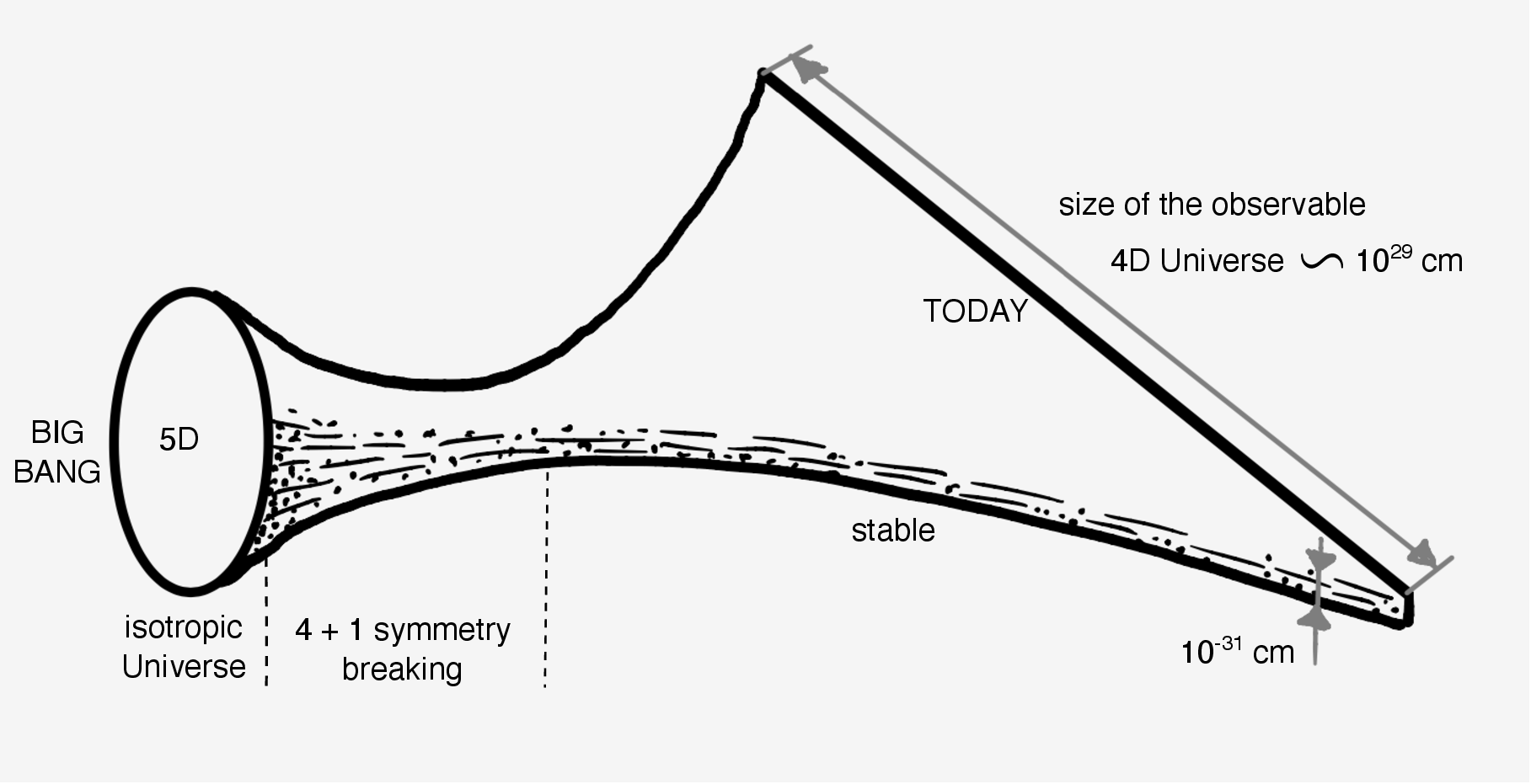}
\caption{$4 + 1$ Symmetry breaking of the Universe}
\label{4 + 1 Symmetry breaking of the Universe}
\end{figure}

We come back now to the Kaluza-Klein group. A good question to ask is: "what is the right place of $\hat{\mathbb{G}}_1$ in the history of the Universe?"

On this basis, we propose a cosmological scenario for the evolution of the structure of elementary particles, inspired by the cosmologies of Kaluza-Klein \cite{Chodos 1980, Sahdev 1984, Matzner 1985, Copeland 1985, Okada 1986, Kerner 1988}. 

On Figure \ref{4 + 1 Symmetry breaking of the Universe}, we represented a brief history of the Universe in 3 episodes:
\begin{itemize}
    \item On the left, near the event called the Big Bang, in the 5D early universe, particles are classified from the momenta of the group $\hat{\mathbb{G}}_1$ and the Universe is isotropic … but unstable.
    \item In the episode 2, the first 3 spatial dimensions expand rapidly while the last one shrinks, 
    \item leading to the 4D era in which today particles are characterized by the momenta of the group $\hat{\mathbb{G}}_0$.
\end{itemize}

\textbf{By this mechanism, the elementary particles can acquire an electric charge as a by-product of the 4 + 1 symmetry breaking of the Universe}. 

\subsection{4 + 1 symmetry breaking (group $\hat{\mathbb{G}}_\omega $)}

During the symmetry breaking, the metric follows the scale change 
$$ \hat{G}_\omega = 
\left[ {{\begin{array}{cc}
       G \hfill &  0 \hfill \\
       0 \hfill &  - \omega^2 \hfill \\
   \end{array} }} \right]  
$$
and its symmetry group $\hat{\mathbb{G}}_\omega $ also. Its elements $\hat{a} = (\hat{C}, \hat{P})$, which are  affine transformations   of $\mathbb{R}^5$,  are similar to those of the Kaluza-Klein group with additional factors $\omega$ in the expressions
$$ \hat{C} = \left[ {{\begin{array}{cc}
       C \hfill  \\
       \xi  \hfill  \\
   \end{array} }} \right], \qquad
\hat{P} = 
\left[ {{\begin{array}{cc}
       P \hfill &  \omega^2 \beta \, P^{*-1} b  \hfill \\
       b^* \hfill &  \beta \hfill \\
   \end{array} }} \right]
$$
where  
\begin{eqnarray}
    b\in \mathbb{R}^4, \quad\qquad 
    \beta = \sqrt{1 + \omega^2 b^* b} 
    \quad\qquad \qquad\qquad\qquad\qquad \qquad\qquad
    \quad\quad \nonumber\\
P = P_L B, \qquad P_L \;\; \mbox{is a Lorentz transformation},
\qquad B = 1_{\mathbb{R}^4} + \frac{\omega^2}{\beta + 1} \, b \, b^* 
\end{eqnarray}
$\hat{\mathbb{G}}_\omega $ is a Lie group of dimension 15. There are two limit cases: for $\omega = 1$, we recover the Kaluza-Klein group and, for the singular case $\omega = 0$, we have already studied the group $\hat{\mathbb{G}}_0$ of Physics today.

\begin{figure}[h!]
\centering
\includegraphics[scale=.60]{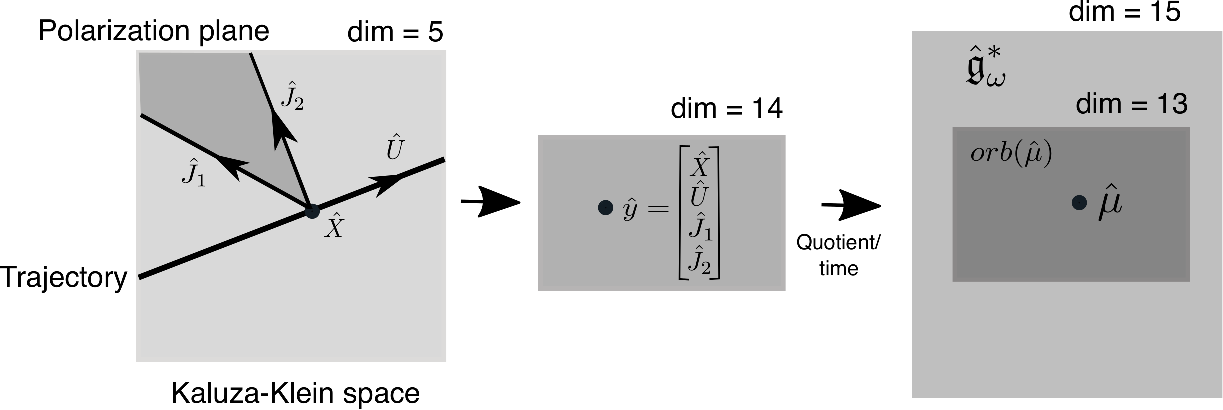}
\caption{Coadjoint orbit of a particle with spin for Kaluza-Klein group}
\label{fig Particle with spin Kaluza-Klein}
\end{figure}

In dimension 5, what is new?

Instead of a polarization line in 4D, there is a polarization plane defined by an orthonormal basis $\hat{J}_1$ and $\hat{J}_2$ (Figure \ref{fig Particle with spin Kaluza-Klein}). We skip the details because the reasoning is very similar to the one with Poincaré's group. Calculating the dimension of the orbit, we find that it is defined by two independent invariants, taken to be the rest mass 
$$ \quad m_0 = \sqrt{\hat{\Pi}^* \hat{\Pi}}
$$
and the spin
$$ s  = \frac{\sqrt{- (\mbox{pol}_{\hat{\mu}}  (\hat{J}_1))^* \mbox{pol}_{\hat{\mu}}  (\hat{J}_1)}}{\sqrt{\hat{\Pi}^*\hat{\Pi}} } 
$$
Let us remark that this is a nice expression for the spin in which the polarization map $\mbox{pol}_{\hat{\mu}}$ appears, a surjective linear map from the 5D space into the polarization plane.
Then if we span $\mathbb{R}^5$, we generate the polarization plane. Next we pick up an orthonormal basis $(\hat{J}_1, \hat{J}_2)$ and we calculate the value of the spin.

\section{The pullback connection}

If there is a Lie group $G$, there is a $G$-structure, then an associated connection. For $\hat{\mathbb{G}}_0$, we will build it by a pullback. We expect to find the Lorentz force in the equation of motion of a particle.

\subsection{$\hat{\mathbb{G}}_0$-connection}

\begin{figure}[h!]
\centering
\includegraphics[scale=.80]{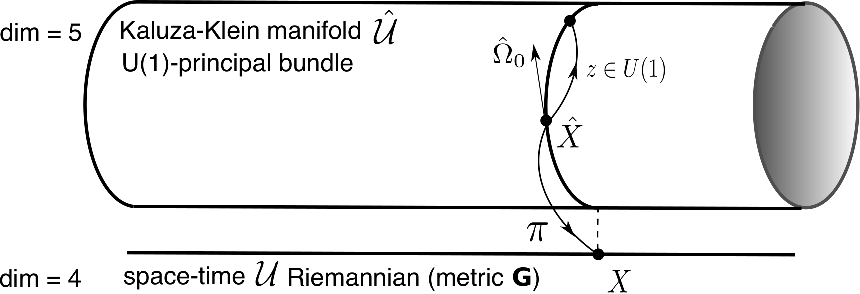}
\caption{Kaluza-Klein manifold}
\label{fig Kaluza-Klein manifold}
\end{figure}

Our goal is to build an Ehresmann connection on the fiber bundle of the frames of $\hat{\mathbb{G}}_0$ resulting from five hypotheses of which the first three are (Figure \ref{fig Kaluza-Klein manifold}):
\begin{itemize}
    \item  \textbf{[H1]} The spacetime, endowed with the metric $\bm{G}$, is Riemannian.
    \item \textbf{[H2]} The Kaluza-Klein 5D manifold $\hat{\mathcal{U}}$ is a principal bundle of structure group the torus $U(1)$, projection $\pi$ and base the spacetime, endowed with:
     \begin{itemize}
         \item the pullback of the metric $\hat{\bm{G}}_0 = \pi^* \bm{G}$ (which is not a metric)
         \item the vector field $\hat{\bm{\Omega}}_0$ whose the integral curves are the fibers
     \end{itemize}
     A $\hat{\mathbb{G}}_0$-basis $(\hat{\bm{e}}'_\alpha)$, is a basis in which $\hat{\bm{G}_0}$ and $\hat{\bm{\Omega}}_0$ are represented by the $5 \times 5$ and $5 \times 1$ invariant matrices
$$        \hat{G}'_0 = \left[ {{\begin{array}{cc}
       G'_0 \hfill &  0 \hfill \\
       0 \hfill &  0 \hfill \\
   \end{array} }} \right], \qquad 
 \hat{\Omega}'_0 =
    \left[ {{\begin{array}{cc}
       0 \hfill  \\
       1  \hfill  \\
   \end{array} }} \right] \qquad \mbox{with the Minkowski metric} \, G'_0
$$
    \item \textbf{[H3]} A $\hat{\mathbb{G}}_0$-\textbf{connection} $\hat{\nabla}$ is a field  of 1-form field on the fiber bundle of $\hat{\mathbb{G}}_0$-bases with values in the Lie algebra of the group
    $$ (\hat{\bm{e}}'_\alpha) \mapsto \hat{\Gamma}'\in \hat{\mathfrak{g}}_0
    $$
    Therefore, considering an infinitesimal perturbation of the identity, we obtain
    $$     \hat{\Gamma}' =
        \delta \left[ \begin{array}{cc}
       P \hfill &  0 \hfill \\
       b^* \hfill &  1 \hfill \\
   \end{array}  \right] =
        \left[ \begin{array}{cc}
       \delta P \hfill &  0 \hfill \\
       \delta b^* \hfill &  0 \hfill \\
   \end{array}  \right] =
    \left[ \begin{array}{cc}
       \Gamma' \hfill &  0 \hfill \\
       \Gamma'^5 \hfill &  0 \hfill \\
   \end{array}  \right]
    $$
    where the $4 \times 4$ matrix $\Gamma'$ is skew-adjoint and $\Gamma'^5$ is a 4-row.
\end{itemize}

\subsection{The pullback connection}

\begin{figure}[h!]
\centering
\includegraphics[scale=.75]{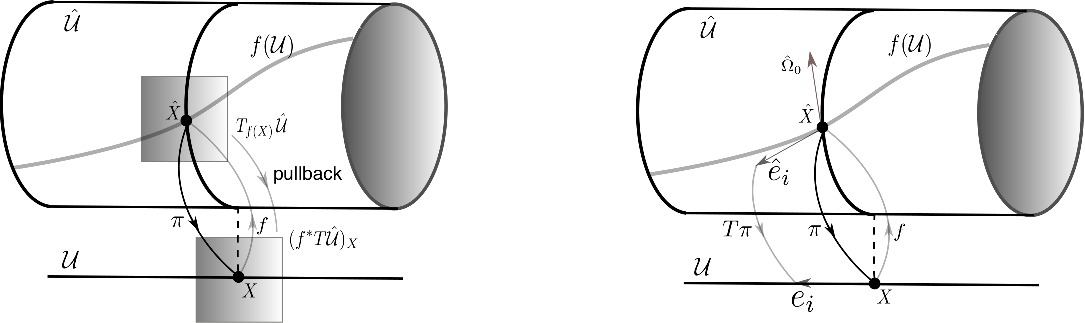}
\caption{Pullback connection}
\label{fig Pullback connection}
\end{figure}

To return to the space-time, one would be tempted to zoom out by the operation of scaling back but the new coordinate $\hat{X}^5 = \omega \hat{X}'^5 $ becomes singular when $\omega$ tends to zero, which leads to a deadlock. To break it, we consider a section $f$ of the principal fiber bundle $\hat{\mathcal{U}}$ and the pullback $f^* T \hat{\mathcal{U}}$ of the tangent bundle by sliding the fiber from $f(X)$ to $X$ (Figure \ref{fig Pullback connection}, on the left). The base space is 4D but the fibers are 5D. Then we introduce a new hypothesis:
\begin{itemize}
    \item \textbf{[H4]} The space-time $\mathcal{U}$ being endowed with the \textbf{pullback connection} $f^* \hat{\nabla}$, we can differentiate 
    only with respect to a vector $\bm{U}$ tangent to the space-time
    $$  (f^* \hat{\nabla})_{\bm{U}} (f^* \hat{\bm{W}}) = f^* (\hat{\nabla}_{(Tf)\bm{U}} \hat{\bm{W}}) \qquad
   \bm{U} \; \mbox{tangent to }\; \mathcal{U}, \qquad
   \hat{\bm{W}} \; \mbox{tangent to} \; \hat{\mathcal{U}} 
    $$
    We impose that it is \textbf{torsion-free}
    $$ \hat{\nabla}_{(Tf) \, \bm{U}} (Tf) \bm{V}
    - \hat{\nabla}_{(Tf) \, \bm{V}} (Tf) \bm{U}
    - (Tf) \,\lbrack \bm{U}, \bm{V}\rbrack 
    = \bm{0}
    $$
    By convention, the Latin indices run from 1 to 4 and the Greek ones from 1 to 5.
    If we work in a basis adapted to the section $f$, \textit{i.e.} whose first 4 vectors $\hat{\bm{e}}_i$ are tangent to $f (\mathcal{U} ) $ and the fifth is tangent to the fiber (Figure \ref{fig Pullback connection}, on the right)
    $$  (\hat{\bm{e}}_1, \cdot \cdot,\hat{\bm{e}}_4, \hat{\bm{\Omega}}_0) 
    $$
    and the basis $(\bm{e}_i)$ of the space-time obtained by projection of the $\hat{\bm{e}}_i$
    $$ \bm{e}_i  = (T \pi) \, \hat{\bm{e}}_i 
    $$
    the torsion-free condition leads to the 50 scalar relations 
    $$ \Gamma^\mu_{ij} 
         - \Gamma^\mu_{ji} - \delta^\mu_k c^k_{ij} 
         = 0
    $$
    where the contravariant indices can take values from 1 to 5, the covariant indices from 1 to 4, and the $c^k_{ij}$ are the structure coefficients of the frame. 
\end{itemize}

\subsection{Equation of motion of a charged particle}

\begin{itemize}
    \item  \textbf{[H5]} we claim that the linear 5-momentum $\hat{\Pi}$ represents a 1-form $\hat{\bm{\Pi}}$ on $\hat{\mathcal{U}}$ and that the motion of a particle and the evolution of the electric charge is such that $\hat{\bm{\Pi}}$ is \textbf{parallel-transported}
    $$  (f^* \hat{\nabla})_{\bm{U}} (f^* \hat{\bm{\Pi}}) = \bm{0}
    $$
    \item \textbf{Integrability condition}. Until now, we were working in a $\hat{\mathbb{G}}_0$-basis whose projection on the space-time is a non-integrable basis (what is called a "moving frame")
    $$ (\hat{\bm{e}}'_\alpha)  = ( \hat{\bm{e}}'_i,\hat{\Omega}_0) 
    \quad \overset{T \pi}{\longrightarrow} \quad 
    (\bm{e}'_i) = ((T \pi) \hat{\bm{e}}'_i)
    $$
    Then we use a $5 \times 5$ transformation matrix $\hat{P}$ 
    $$ \hat{P} =
    \left[ \begin{array}{cc}
       P \hfill &  0 \hfill \\
       - 2 \, A^* \hfill &  1 \hfill \\
   \end{array} \right]
    $$
    broken down by blocks where $A$ will turn out to be the \textbf{electromagnetic 4-potential}, to work in an integrable basis associated with coordinates $(X^i)$ (called also a "natural frame")
    $$ (\hat{\bm{e}}_\alpha)  
    = ( \hat{\bm{e}}_i,\hat{\Omega}_0) \quad 
    \overset{T \pi}{\longrightarrow} \quad 
    (\bm{\partial}_i) = ((T \pi) \hat{\bm{e}}_i)$$
    In this new basis, the calculation shows that the connection matrix is of this form
    $$ \hat{\Gamma} =
    \left[ \begin{array}{cc}
       \Gamma \hfill &  0 \\
       \Gamma^5 \hfill & 0 \hfill \\
   \end{array}  \right]
    $$
    \item Using the free-torsion condition  with the contravariant index running from 1 to 4
    $$ \Gamma^k_{ij} =  \Gamma^k_{ji} 
    $$
    we demonstrate that the sub-connection $\Gamma$ is that of Levi-Civita (this is the revisited  fundamental lemma of the Riemannian geometry). Then we rewrite $\hat{\Pi} $ in the new basis 
    $$ \hat{\Pi}^* 
    =  \hat{\Pi}'^*\hat{P} 
    = \left[ \Pi'^*, q' \right] \, \left[ \begin{array}{cc}
       P \hfill &  0 \hfill \\
       - 2 \, A^* \hfill &  1 \hfill \\
   \end{array} \right]
    =  \left[ \Pi^* - 2 \, q \, A^*, q\right]
    $$
    and, using  hypothesis \textbf{[H5]}, we prove that the rest mass $m_0$ and the electric charge $q$ are --as expected-- \textbf{integrals of the motion} and that the components of the row $\Gamma^5$ are of the following form
    $$  \Gamma^5_j = U^k F_{kj} - 2 \, \nabla_U A_j
    $$
    where $F_{kj}$ is skew-symmetric. Finally, we use again the free-torsion condition but with the contravariant index equal to 5, and we show that $F_{ij}$ is the \textbf{electromagnetic field}
    $$ F_{ij} = \partial_i A_j -  \partial_j A_i  
    $$
    Finally the equation of the motion takes the expected form with the \textbf{Lorentz force} in the right hand member
    $$ m_0 U^k \nabla_k U^i = - q \, F^i_j U^j 
    $$
    \item It can also be shown that a change of section $f$ corresponds to the classic gauge transformation. 
\end{itemize}

\section{The variational relativity}

We come to the third part: we shall be going to extend the variational relativity. We expect to find the second group of Maxwell’s equations, the first group being equivalent to the fact that $F$ is the exterior derivative of $A$.

\subsection{Stationary action principle}

We use Palatini's variational approach, \textit{i.e.} the connection and the curvature are independent variables. Inspired by the presentation given in \cite{GR}, we start from the following principles:
\begin{itemize}
    \item [{\bf (P1)}] To every {\bf physical phenomenon} corresponds a field $z$  and a Lagrangian $ L_{(z)} (z, \partial_i z, G, A^*)$
     \item [{\bf (P2)}] The \textbf{action} 
     $$ S = \int_{\mathcal{D}} L \, vol
           = \sum_z \int_{\mathcal{D}} L_{(z)} \, vol
     $$
     over a domain  $\mathcal{D}$ of the space-time $\mathcal{U}$
     is stationary for every variation of the metric $G$, the electromagnetic potential $A$ and the fields $z$, which leads to the definition of the tensor $T_{(z)}$  , the vector $\tilde{T}_{(z)}$  and the variational derivatives $W_{(z)}$
     $$  T_{(z)} = 2 \, \partial_G L_{(z)} + L_{(z)} G^{-1}, \quad
     \tilde{T}_{(z)} = \partial_{A^*} L_{(z)}, \quad\quad\;\;\;
     W_{(z)} = \partial_z L_{(z)} - \partial_i (\partial_{\partial_i z} L_{(z)})
     $$
     hence the stationarity condition, composed of the Einstein equation, the Euler-Lagrange equations of each phenomenon and an additional equation associated with the electromagnetic potential A.
     $$  T = \sum_z T_{(z)}  = 0, \qquad
     \tilde{T} = \sum_z \tilde{T}_{(z)} = 0, \qquad
     W_{(z)}  = 0
     $$
     The two former conditions are called \textbf{field equations}.
\end{itemize}

We can then demonstrate \textbf{conservation laws} whose first contains the electromagnetic field $F$
$$ \mbox{div} \, T_{(z)}  + \tilde{T}_{(z)}  \cdot F = 0, \qquad
\mbox{div} \,  \tilde{T}_{(z)} = 0
$$

\subsection{The matter and its motion}

The first phenomenon considered is the matter. A particle is located by its Lagrangian coordinates as a function of a space-time event which gives the equation of its trajectory \cite{GR, Soper 1976}
$$ a = \pi_A (X)
$$
The Lagrangians $L_{(z)}$  must be invariant by any space-time diffeomorphism. To satisfy this condition, the Lagrangian of the matter 
$$ L_M = L_M \left( a, \frac{\partial a}{\partial  X}, G, A^*\right) 
$$
must depend on derivatives of $a$ via the conformation tensor
$$ H = \frac{\partial a}{\partial  X} \, G^{-1} \frac{\partial a}{\partial  X}^T
$$
For a fluid, the Lagrangian depends on $h = - \det H$
$$ L_M = L_M \left( a, h, G, A^*\right) 
$$
We then define the mass density and the electric charge density
$$ \rho_m = \rho_{m0} (a) \, \sqrt{h}, \qquad
\rho_e = \rho_{e0} (a) \, \sqrt{h}
$$

In its simplest form, the Lagrangian of the matter is equal to 
$$ L_M = \kappa \, \left[ L_m (a, h, G) + L_e (a, h, A^*) \right]
$$
where $\kappa$ is a coupling constant. The contribution of the mass is 
$$ L_m (a, h, G) = \rho_{m0} (a) \, \lbrack \sqrt{h} + \psi (h) \rbrack 
$$
where $\psi$ is the internal energy. The Lagrangian of the electric charge is
$$ \L_e (a, h, A^*)  = - \rho_{e0} (a) \, \sqrt{h}  \, A^* U
$$
Define the energy density $\rho$ and the pressure $p$ by 
$$ \rho = L_M / \kappa, \qquad
    p  = \rho_{m0} \, (2 \, h \, \partial_h \psi - \psi)
$$
one obtains the \textbf{conservation law of the linear momentum} 
$$ \nabla_i \lbrack (\rho + p) U^i U_j - p \, \delta^i_j \rbrack
     - \rho_e U^k F_{kj} = 0
$$
where the last term is the Lorentz force, and the \textbf{law of conservation of the electric current} 
$$ \nabla_i (\rho_e U^i) = 0
$$

\subsection{Field equations the gravitation and electromagnetism}

We now consider the phenomena of gravitation and electromagnetism, then the pullback connection. The Lagrangian depends on its derivatives via the curvature tensor of the space-time 
$$      \hat{\nabla}_{(Tf) \, \bm{U}} \, (\hat{\nabla}_{(Tf) \, \bm{V}} \, \hat{\bm{W}})
   - \hat{\nabla}_{(Tf) \, \bm{V}} \, (\hat{\nabla}_{(Tf) \, \bm{U}}
      \, \hat{\bm{W}})
   - \hat{\nabla}_{(Tf) \,\lbrack \bm{U}, \bm{V}\rbrack} \, \hat{\bm{W}}
   = \hat{\bm{R}} (\bm{U}, \bm{V}) \, \hat{\bm{W}}
$$
The non-zero components are the ones with 4 space-time Latin indices 
$$ R^p_{ijk} = \hat{R}^p_{ijk} 
 =  \Gamma^p_{i\mu} \Gamma^\mu_{jk}
  - \Gamma^p_{j\mu} \Gamma^\mu_{ik}
  + \partial_i \Gamma^p_{jk}
  - \partial_j \Gamma^p_{ik}
$$
and the ones with 3 Latin covariant indices and the contravariant index equal to 5
$$ \tilde{R}_{ijk} = \hat{R}^5_{ijk} 
 =  \nabla_k F_{ji} + 2 A_q R^q_{ijk}
$$
in which the Maxwell equations are glimpsed, provided that the indices $i$ and $j$ are raised and we perform a contraction on the indices $k$ and $j$. This is done by adopting the following Lagrangian 
$$ L_G = L_G (\hat{\Gamma},  \partial_i \hat{\Gamma}, G, A^*) = L_G (\hat{\Gamma},  \hat{R}, G, A^*)
$$
and assuming once again a decomposition
$$  L_G = L_g (R, G)  + L_{em} (\tilde{R}, G, A^*) 
 = - \Lambda 
   + \frac{1}{2} \, G^{ij} R_{ij} 
   - \tilde{k}\, A_r G^{ri} G^{jk} \tilde{R}_{ijk} 
$$
where $\Lambda$ is the cosmological constant, $R_{ij}$ is the Ricci tensor and $\tilde{k}$ is a new coupling constant. 

The stationarity condition gives 14 field equations for 14 unknown potentials $G_{ij}, A_i$ 
$$       R^{ij} - \frac{1}{2} \, R \, G^{ij} + \Lambda \, G^{ij} 
    - \tilde{k} \, \lbrack A^{( i} \, \tilde{R}^{j)\;\,k}_{\;\;\;k}  
                           + A_r \tilde{R}^{r(ij)} 
    - \frac{1}{2} \, \tilde{R} \, G^{ij}        \rbrack 
= \kappa \, \lbrack   (\rho + p) U^i U^j - p \, G^{ij} \rbrack 
$$
$$   - \tilde{k} \, \lbrack \nabla_j F^{ji} + 2 A_q R^{qij}_{\quad j} \rbrack
    = \kappa \, \rho_e U^i  
$$

\subsection{Classical limit}

In the absence of electromagnetism, we find Einstein's equations and, at the Newtonian approximation, we obtain, by identification with Newton's law of universal attraction, the first coupling constant $\kappa$ in terms of the gravitation constant $G_N$ (with the $N$ in tribute to Newton)
$$ \kappa = 8 \, \pi \, G_N
$$
We are now interested in the second equation. With the Galilean gravitation of components the gravity $\Gamma^a_{tt} = - g^a$ and the Coriolis effect $\Gamma^a_{tb} = \Gamma^a_{bt} = \Omega^a_b$, an astonishing simplification occurs: $\Omega_{ab}$ and $F^{ij}$ being skew-symmetric, the covariant divergence of $F$ is equal to its simple divergence (as if the space-time was flat)
$$      \nabla_j F^{ji} 
   = \partial_j F^{ji} 
$$
Under laboratory conditions, $g$ and $\Omega$ are constant, which, by identification with Coulomb’s law, gives the second coupling constant $\tilde{k}$ as a function of $G_N$ and the permittivity $\epsilon_0$ 
$$ \tilde{k} =  8 \, \pi \, G_N \, \epsilon_0
$$
and we find the second group of Maxwell’s equations
$$ \partial_j F^{ji}  =  - \frac{1}{\epsilon_0} \, \rho_e U^i 
$$
In the first equation, the coupling term is of the order of the stored electrical energy density, very negligible compared to the energy density $\rho$ ($\rho c^2$ in international units). 

The very weak coupling between gravitation and electromagnetism can be related to Dirac's Large Number Hypothesis. The constant 
$$ \tilde{k} = \frac{2 \, e^2}{m_e m_p} \, \frac{F_g}{F_e}
$$
is proportional to the ratio of the gravity force to the electric force between an electron and a proton, a dimensionless number of the order of $10^{-40}$. It is interesting to note that it is equal to the square of the ratio between the Klein's cylinder radius 
$$  l_K = 8 \, \pi^{3/2} \, \frac{h \, \sqrt{G_N \epsilon_0}}{e \, c} = 0.238 \times 10^{-31} \, \mbox{cm}
$$
and twice the geometric mean of the Compton wavelengths of the electron and the proton 
$$  \frac{F_g}{F_e} = \left( \frac{l_K} {2 \, \sqrt{\lambda_C \lambda_{C, p}}}\right)^2
$$

\section{Conclusions and perspectives}

\begin{itemize}
\item We proposed a symmetry group $\hat{\mathbb{G}}_0$ for which the electric charge of a particle is an \textbf{invariant},
  a $\hat{\mathbb{G}}_0$-connection allowing recovery of the expected equation of motion of a particle today, including  the \textbf{Lorentz force}, 
  and a 5D extension of the relativity allowing recovery  of the \textbf{Maxwell equations} in the classical limit.
\item $\hat{\mathbb{G}}_0$ \textbf{is the symmetry group of the electrodynamics 
         compatible with observations today}.
\item  The symmetry group $\hat{\mathbb{G}}_1$ of the Kaluza-Klein theory leads to 
 \textbf{a unified theory merging the gravitational and electromagnetic forces, 
  relevant in the early Universe}.
\end{itemize}

The sound idea of a fifth dimension, if used in a suitable geometric formalism taking into account its overwhelmingly small size, seems now more than ever to offer a promising future for new developments, among which we emphasize:

\begin{itemize}
    \item[1.] 
    For the characterization of charged elementary particles, we largely relied on \cite{SSD, SSDEng} without utilizing the full potential of the theoretical framework, in particular the geometric quantization. These issues have not been considered yet in this work but would deserve to be investigated.
    \item[2.] 
    The next step would be to extend Kaluza-Klein theory to a non-abelian gauge group as in \cite{Kerner 1968} with the mathematical tools developed in the present paper, in particular the coadjoint orbit method.
    \item[3.]
    Going back in time, it would be worth revisiting works on early universe cosmology that we think they did not receive the welcome they deserved because based on Kaluza-Klein theory whose weaknesses were known. They could provide a better understanding of Dirac's large number hypothesis as suggested in \cite{Chodos 1980}  and to offer a resolution to the horizon problem as claimed in \cite{Sahdev 1984}.   
\end{itemize}

\vspace{0.3cm}

\textbf{Acknowledgements}

\vspace{0.3cm}

The author would like to thank Leonid Ryvkin for discussions that help to clarify certain points and to Richard Kerner whose advice was very valuable to me. He is also grateful to Claude Vall\'ee for drawing the author's attention on the reference \cite{Cartan 1934}.


\end{document}